\documentclass[review,12pt]{elsarticle}
\usepackage{amssymb}
\usepackage{graphicx}
\usepackage{color}

\journal{Physica A}

\begin{document}

\begin{frontmatter}

\title{Granular gas of viscoelastic particles in a homogeneous cooling state.}

\author[msu]{A. S. Bodrova\corref{cor1}}
\ead{bodrova@polly.phys.msu.ru}
\author[lei,msu]{N. V. Brilliantov}

\cortext[cor1]{Corresponding author}

\address[msu]{Faculty of Physics,
M.V.Lomonosov Moscow State University,
Leninskie Gory,\\ Moscow, 119991, Russia}
\address[lei]{Department of Mathematics,
University of Leicester
University Road,\\
Leicester LE1 7RH, UK}

\begin{abstract}
Kinetic properties of a  granular gas of viscoelastic particles in a homogeneous cooling state are
studied analytically and numerically. We employ the most recent expression for the velocity-dependent
restitution coefficient for colliding viscoelastic particles, which allows to describe systems with
large inelasticity. In contrast to previous studies, the third coefficient $a_3$ of the Sonine
polynomials expansion of the velocity distribution function is taken into account. We observe a
complicated evolution of this coefficient. Moreover, we find that $a_3$ is always of the same order of
magnitude as the leading second Sonine coefficient $a_2$; this contradicts the existing hypothesis that
the subsequent Sonine coefficients $a_2$, $a_3 \ldots$,  are of an ascending order of a small parameter,
characterizing particles inelasticity. We analyze evolution of the high-energy tail of the velocity
distribution function. In particular, we study the time dependence of the tail amplitude and of the
threshold velocity, which demarcates the main part of the velocity distribution and the high-energy
part. We also study  evolution of the self-diffusion coefficient $D$ and explore the impact of the third
Sonine coefficient on the self-diffusion. Our analytical predictions  for the third Sonine coefficient,
threshold velocity and the self-diffusion coefficient are in a good agreement with the numerical
finding.

\end{abstract}


\end{frontmatter}

\section{Introduction}
 An ensemble of macroscopic particles, which move ballistically
between dissipative collisions,  is usually termed as a {\it granular gas}, e.g.
\cite{book,Goldhirsh_AnRev2003,hinrichsen, book1, book2} and the loss of energy at impacts is quantified
by a  restitution coefficient $\varepsilon$,
\begin{equation}
\label{eq:def_eps} \varepsilon=\left|\frac{\left(\vec{v}_{12}^{\, \prime} \cdot
\vec{e}\right)}{\left(\vec{v}_{12}\cdot \vec{e}\right)}\right| \, .
\end{equation}
Here $\vec{v}_{12}$ and $\vec{v}_{12}^{\, \prime}$ are relative velocities of two particles before and
after an impact and $\vec{e}$ is the unit vector, connecting their centers at the collision instant. In
what follows we assume that the particles are smooth spheres  of diameter $\sigma$, that is, we do not
consider the tangential motion of particles. Then the restitution coefficient yields the after-collision
velocities of particles in terms of the pre-collision ones,
\begin{equation}
\label{eq:v1v2aft_pre} \vec{v}_{1}^{\prime} = \vec{v}_{1} - \frac12(1+\varepsilon) \left(\vec{v}_{12}
\cdot \vec{e}\, \right) \vec{e} \, , \qquad \quad \vec{v}_{2}^{\prime} = \vec{v}_{2} +
\frac12(1+\varepsilon) \left(\vec{v}_{12} \cdot \vec{e}\, \right) \vec{e} \, .
\end{equation}
The simplest model for the restitution coefficient, $\varepsilon ={\rm const}$, facilitates
significantly the theoretical analysis and often leads to qualitatively valid results, e.g.
\cite{book,Goldhirsh_AnRev2003}.  This assumption, however, contradicts the experimental observations,
e.g. \cite{goldsmit,bridges,kuwabara} and basic mechanical laws, e.g.  \cite{tanaka,ramirez}. A
first-principle analysis of a dissipative collision may be performed, leading to a conclusion that
$\varepsilon$ must depend on the relative velocity of the colliding particles
\cite{kuwabara,bril96,Morgado97}. If the impact velocity is not very high (to avoid plastic deformation
of particles material) and not too small (to avoid adhesive interactions at a collision
\cite{bril2007}), the viscoelastic contact model may be applied \cite{bril96}. This yields the
velocity-dependent restitution coefficient \cite{schwager1998,ramirez}:
\begin{equation}
\varepsilon=1- C_{1}\delta\left(2u(t)\right)^{1/10}\left|\left(\vec{c}_{12}\vec{e} \,
\right)\right|^{1/5}+ C_{2}\delta^{2}\left(2u(t)\right)^{1/5}\left|\left(\vec{c}_{12}\vec{e} \,
\right)\right|^{2/5} \pm \ldots \label{eps}
\end{equation}
Here $C_{1}\simeq 1.15$, $C_{2} \simeq 0.798$ and  $u(t)=T(t)/T(0)$  is the dimensionless temperature,
expressed in terms of current granular temperature, \label{temp}
\begin{equation}
\frac{3}{2}nT(t)=\int d\vec{v}\frac{m\vec{v}^{\, 2}}{2}f(\vec{v},t) = \frac{m v_{T}(t)}{2}\, ,
\end{equation}
with $f(\vec{v},t)$ being the velocity distribution function of grains, $n$ is a gas number density, $m$
is the particle mass  and $v_T(t)$ is the thermal velocity;  $\vec{c}_{12}=\vec{v}_{12}/v_{T}$ is,
correspondingly, the dimensionless impact velocity.

The restitution coefficient $\varepsilon$ (\ref{eps}) depends on the small dissipation parameter
$\delta$
\begin{equation}
\delta=A\kappa^{\frac{2}{5}}\left(\frac{T(0)}{m}\right)^{\frac{1}{10}}\, ,
\end{equation}
proportional to the dissipative constant $A$,
\begin{equation}
A=\frac{1}{3}\frac{\left(3\eta_2-\eta_1\right)^2}{\left(3\eta_2+2\eta_1\right)}\frac{(1-\nu^2)(1-2\nu)}{Y\nu^2}\,,
\end{equation}
which depends on the Young modulus of the particle's material $Y$, its Poisson ratio $\nu$ and the
viscous constants $\eta_1$ and $\eta_2$, see e.g. \cite{book}. The parameter $\kappa$ reads:
\begin{equation}
\kappa=\left(\frac{3}{2}\right)^{\frac{3}{2}}\frac{Y\sqrt{\sigma}}{m\left(1-\nu^2\right)} \,.
\end{equation}

Due to inelastic nature of inter-particle collisions, the velocity distribution function of grains
deviates from the Maxwellian, so that the dimensionless distribution function \cite{esipov}
\begin{equation}
f(\vec{v},t)=\frac{n}{v_{T}^{3}}\tilde{f}(\vec{c},t) \, , \label{scaled}
\end{equation}
may be represented in terms of the Sonine polynomials expansion, e.g. \cite{goldshtein,vannoije,book,pf}:
\begin{equation}
\tilde{f}(\vec{c},t)=\phi(c)\left(1+\sum_{p=1}^{\infty}a_{p}(t)S_{p}(c^2)\right), \label{sonine}
\end{equation}
where $\phi(c)=\pi^{-3/2}\exp(-c^{2})$ is the dimensionless Maxwellian distribution and the first few
Sonine polynomials read \cite{book},
\begin{eqnarray}
\label{eq:Sonine}
&& S_0(x) =1 \, ,  \quad S_1(x)=\frac32 -x \, , \\
&& S_2 (x) =\frac{15}{8}-\frac{5}{2} x +\frac{1}{2} x^{2}  \,  ,\\
&& S_3 (x) = \frac{35}{16}-\frac{35}{8}x+\frac{7}{4}x^{2}-\frac{1}{6}x^{3} \, .
\end{eqnarray}
Provided that the expansion (\ref{sonine}) converges, the Sonine coefficients $a_{p}$ completely
determine the form of $\tilde{f}(\vec{c},t)$. According to the definition of temperature $a_{1}(t)=0$
\cite{book}, that is, the first non-trivial coefficient in the expansion (\ref{sonine}) is $a_{2}(t)$.
For the case of a constant $\varepsilon$ the coefficient $a_2$ has been found analytically
\cite{goldshtein,vannoije} and the coefficient $a_3$ analytically \cite{breakdown} and numerically
\cite{breakdown,Brey1996,Santos1999,Nakanishi2003}. The expansion (\ref{sonine}) quantifies the
deviation of the distribution function from the Maxwellian for the main part of the velocity
distribution, that is, for $c \sim 1$;  the high-velocity tail of $\tilde{f}(\vec{c}\, ) $ for $c \gg 1$
is exponentially over-populated \cite{esipov,vannoije,book,pf} and requires a separate analysis.

The impact-velocity dependence of the restitution coefficient, as it follows from the realistic
visco-elastic collision model, has a drastic impact on the granular gas properties. Namely, the form of
the velocity distribution function and its time dependence significantly change \cite{vdve}, similarly
changes the time dependence of the kinetic coefficients \cite{Bril_PRE2003,Bril_in_NLP}. Moreover, the
global behavior of the system qualitatively alters: Instead of evolving to a highly non-uniform final
state of a rarified gas and dense clusters, as predicted for $\varepsilon ={\rm const.}$
\cite{mcnamara,goldhirsch}, the clustering \cite{goldhirsch} and the vortex formation
\cite{BritoErnst98} occurs in a gas of viscoelastic particles only as a transient phenomena
\cite{BrilPRL2004}.

Based on  the restitution coefficient given by Eq. (\ref{eps}) the theory of granular gases of
viscoelastic particles has been developed, e.g.  \cite{book}, where as in the case of a constant
$\varepsilon$, only the first non-trivial Sonine coefficient $a_2(t)$ has been taken into account.
Although the evolution of the high-velocity tail of $\tilde{f}(\vec{c}\, ) $ has been analyzed
\cite{vdve}, neither the location of the tail, nor its amplitude was  quantified.

Recently, however, it has been shown for the case of a constant restitution  coefficient, that the next
Sonine coefficient $a_3$ is of the same order of magnitude as the main coefficient $a_2$
\cite{breakdown}; moreover, it was also shown that the amplitude of the high-velocity tail of
$\tilde{f}(\vec{c}\,) $  and its contribution to the kinetic coefficients may be described
quantitatively \cite{vdeps,tail2007}.

Finally, a new expression for the velocity-dependent restitution coefficient $\varepsilon$ has been
derived \cite{delayed}, which takes into account the effect of  "delayed recovery"  in  a collision. The
delayed recovery implies,  that at the very end of an impact, when the colliding particles have already
lost their contact, their material remains deformed \cite{delayed}. This affects the total dissipation
at a collision, so that the  revised $\varepsilon$ reads \cite{delayed}:
\begin{equation}
\varepsilon=1+\sum_{i=1}^{\infty}h_{i}\delta^{i/2}\left(2u(t)\right)^{i/20}
\left|\left(\vec{c}_{12}\vec{e}\right)\right|^{i/10} \, . \label{eps1}
\end{equation}
Here $h_{1}=0$, $h_{2}=-C_{1}$, $h_{3}=0$,  $h_{4}=C_{2}$ and the other numerical coefficients up to
$i=20$ are given in \cite{delayed}. Fig. \ref{Gu} (inset) illustrates the
dependence of the restitution coefficient (\ref{eps1}) on the dissipative parameter $\delta$ for $u=1$
for collisions with the  characteristic thermal velocity, that is for $\left|\left(\vec{c}_{12} \cdot
\vec{e}\right)\right|=1$.

In the present study we use the revised expression (\ref{eps1}) for  $\varepsilon$  and investigate the
evolution of the velocity distribution function in a gas of viscoelastic particles in a homogeneous
cooling state. With the new restitution coefficient we are able to describe collisions with
significantly larger dissipation, than it was possible before with the previously available expression
for $\varepsilon$. For the larger inelasticity one expects the increasing importance of the next-order
terms in the Sonine polynomials expansion and of the high-energy tail of the velocity distribution. In
what follows we study analytically and numerically time evolution of $\tilde{f}(\vec{c},t) $, using the
Sonine expansion up to the third-order term and analyze the amplitude and slope of the high-energy tail
of $\tilde{f}(\vec{c},t) $. In addition,  we consider self-diffusion -- the only non-trivial transport
process in the homogeneous cooling state and compute the respective coefficient.

The rest of the paper is organized as follows. In the next Sec. II we address the Sonine polynomial
expansion and calculate the time-dependent Sonine coefficients, along with the  granular temperature. In
Sec. III the high-energy tail is analyzed, while Sec. IV is devoted to the self-diffusion coefficient.
Finally, in Sec. V we summarize our findings.

\section{Sonine polynomial expansion: evolution of the expansion coefficients}

Evolution of the velocity distribution function of a granular gas of spherical particles of diameter
$\sigma$ in a homogeneous cooling state obeys the Boltzmann-Enskog equation, e.g.
\cite{book,Goldhirsh_AnRev2003}:

\begin{equation}
\frac{\partial f(\vec{v},t)}{\partial t}=g_{2}(\sigma)I(f,f) \,, \label{be}
\end{equation}
where $I(f,f)$ is the collision integral. Generally, $I$  depends on the two-particle distribution
function $f_2(\vec{v}_1,\vec{v}_2, \vec{r}_{12},t)$. Within  the hypothesis of molecular chaos,
$f_2(\vec{v}_1,\vec{v}_2, \vec{r}_{12},t)=g_{2}(\sigma)f(\vec{v}_1,t) f(\vec{v}_2,t)$ (see e.g.
\cite{book}), the closed-form equation (\ref{be}) for $f(\vec{v},t)$ is obtained;  $g_{2}(\sigma)$
denotes  here a pair correlation function at a contact, which accounts for the increasing collision
frequency due to the effect of excluded volume \cite{book}.

Eq. (\ref{be}) yields the equation for the dimensionless distribution function
$\tilde{f}(\vec{c},t)$ \cite{book},
\begin{equation}
\frac{\mu_{2}}{3}\left(3+c\frac{\partial}{\partial c}\right)\tilde{f}
(\vec{c},t)+B^{-1}\frac{\partial}{\partial t}\tilde{f}(\vec{c},t)=\tilde{I}(\tilde{f},\tilde{f})\, ,
\label{mu2}
\end{equation}
where $B=v_{T}g_{2}(\sigma)\sigma^{2}n$ and
\begin{equation}
\mu_{p}=-\int d\vec{c}c^{p}\tilde{I}(\tilde{f},\tilde{f}),
\label{mup}
\end{equation}
is the $p$-th moment of the dimensionless collision integral:
\begin{equation}
\tilde{I}(\tilde{f},\tilde{f})=\int d\vec{c}_{2} \int d\vec{e} \, \Theta (-\vec{c}_{12}\cdot \vec{e}\, )
\left|-\vec{c}_{12} \cdot \vec{e}\, \right| \left(\frac{1}{\varepsilon^{2}}\tilde{f}(\vec{c}_{1}^{\
\prime\prime},t)\tilde{f} (\vec{c}_{2}^{\
\prime\prime},t)-\tilde{f}(\vec{c}_{1},t)\tilde{f}(\vec{c}_{2},t)\right) \, .\label{II}
\end{equation}
$\Theta(x)$ in the above equation is the Heaviside step-function and the dimensionless velocities
$\vec{c}_{1}^{\, \prime\prime}$ and  $\vec{c}_{2}^{\, \prime\prime}$ are the pre-collision velocities in
the so-called inverse collision, which results with $\vec{c}_{1}$ and $\vec{c}_{2}$ as the
after-collision velocities, e.g. \cite{book}.

Eq. (\ref{mu2}) is coupled to the equation for  the granular temperature, e.g. \cite{book}:
\begin{equation} \label{eq:dTdt} \frac{dT}{dt}=-\frac{2}{3}BT\mu_{2} = -\zeta T \, ,
\end{equation}
which defines the cooling coefficient $\zeta= (2/3)B \mu_2$.

Multiplying Eq. (\ref{mu2}) with  $c^{p}$, integrating over $\vec{c}$ and writing Eq. (\ref{eq:dTdt}) in
the dimensionless form, we obtain for $p=4, \, 6$ \cite{book} (similar equations has been used in
\cite{huthmann}):
\begin{eqnarray}
\cases{\frac{du}{d\tau}=-\frac{\sqrt{2}\mu_{2}}{6\sqrt{\pi}}u^{\frac{3}{2}}\cr
\frac{da_{2}}{d\tau}=\frac{\sqrt{2}\sqrt{u}}{3\sqrt{\pi}}\mu_{2}\left(1+a_{2}\right)-\frac{\sqrt{2}}{15\sqrt{\pi}}\mu_{4}\sqrt{u}\cr
\frac{da_{3}}{d\tau}=\frac{\sqrt{u}}{\sqrt{2\pi}}\mu_{2}(1-a_{2}+a_{3})-\frac{\sqrt{2}}{5\sqrt{\pi}}\mu_{4}\sqrt{u}+
\frac{2\sqrt{2u}}{105\sqrt{\pi}}\mu_{6}}
\label{system}
\end{eqnarray}
Here $\tau=t/\tau_{c}(0)$ is the dimensionless time, measured in the mean collision units,
$\tau_{c}^{-1}(0)=4\sqrt{\pi}g_{2}(\sigma)\sigma^{2}n\sqrt{T(0)/m}$ at initial time $t=0$. The equations
for the Sonine coefficients $a_2$ and $a_3$ in (\ref{system}) are the first two of  the infinite system
of equations for $p=4,\, 6,\, 8, \ldots$ \cite{huthmann,book}. The moments $\mu_2, \, \mu_4, \, \mu_6,
\ldots$ in these equations depend on all the Sonine coefficients $a_k$, with $k=2, \ldots, \infty$, that
is, only the infinite set  of equation is closed. To make the system tractable one has to truncate it.
Following Refs. \cite{breakdown,vestnik} we truncate the Sonine series at the third term, i.e. we
approximate,
\begin{equation}
\tilde{f}(\vec{c},\tau) \simeq \phi(c)\left(1+ a_2(\tau) S_2(c^2) +a_3(\tau) S_3(c^2) \right) \, .
\label{eq:sonine3}
\end{equation}
Within  this approximation $\mu_2, \, \mu_4, \, \mu_6$ in Eqs. (\ref{system}) depend only on $a_2$,
$a_3$ and $u$, which makes the system (\ref{system}) closed.

The moments of the collisional integral $\mu_p$ were calculated analytically up to ${\cal
O}(\delta^{10})$, using the formula manipulation program, explained in detail in \cite{book}. The
complete  expressions are rather cumbersome, therefore, we present here for illustration only linear
approximations  with respect to $a_2$, $a_3$ and $\delta^{\prime}(\tau) = \delta \left( 2 u (\tau)
\right)^{1/10}$:
\begin{eqnarray}
\nonumber &&\mu_2=\omega_0 \left(1+\frac{6}{25}a_2+ \frac{2}{125}a_3\right)\delta^{\prime}(\tau) \\
\label{mu246}
&&\mu_4=\sqrt{2\pi}\left(4a_2-a_3 \right) +7\left(\frac{4}{5} + \frac{129}{125}a_2 -\frac{179}{1250} a_3 \right)\omega_0\delta^{\prime}(\tau)\\
\nonumber &&\mu_6=3\sqrt{2\pi}\left(15a_2- \frac{45}{4}a_3 \right)+3\omega_0 \delta^{\prime}(\tau)
\left( \frac{1079}{100} +\frac{40717}{1250}a_2-\frac{39353}{3125} a_3 \right) \, ,\nonumber
\end{eqnarray}
where $\omega_0=2\sqrt{2\pi}2^{1/10} \Gamma\left(21/10 \right) C_1 \simeq 6.485$.

Since the system of equations (\ref{system}) is strongly non-linear, only a numerical solution is
possible. Still, one can find a perturbative solution in terms of small dissipative parameter $\delta$,
\begin{equation}
a_{2}=a_{20}+a_{21}\delta+..., \qquad a_{3}=a_{30}+a_{31}\delta+..., \qquad
u=u_{0}+u_{1}\delta+...
\end{equation}
The zeroth-order solution (with $a_2(0)=a_3(0)=0$) reads,
\begin{equation}
a_{20}=0 \, ,  \qquad \qquad a_{30}=0\, ,  \qquad \qquad
 u_{0}=\left(1+\tau/\tau_{0}\right)^{-5/3} \, ,
\end{equation}
where  $\tau_{0}^{-1}=2^{6/5}\Gamma\left(21/10\right)C_1\delta/5\simeq0.55\delta$. For the first-order
solution only its  asymptotics for $\tau \to \infty$ may be obtained analytically:
\begin{eqnarray}
&&a_2 = -A_2 \, \delta\left(\tau/\tau_{0}\right)^{-1/6} \quad \quad A_2=
2^{1/5}\frac{157}{500}\Gamma\left(21/10\right)C_1 \simeq 0.44   \label{eq:a2}\\
&&a_3 = -A_3 \, \delta\left(\tau/\tau_{0}\right)^{-1/6} \quad \quad
A_3= 2^{1/5}\frac{28}{500}\Gamma\left(21/10\right)C_1  \simeq 0.08  \label{eq:a3}\\
&&u=\left(\tau/\tau_{0}\right)^{-5/3}+ q \delta\left(\tau/\tau_{0}\right)^{-11/6}, \label{u11}\\
&&q=2^{\frac{1}{5}}C_1\left(\frac{2383}{15625}\Gamma\left(21/10\right)+
\frac{\Gamma\left(16/5\right)}{\Gamma\left(21/10\right)}\right)\simeq 3.28 \, . \nonumber
\end{eqnarray}
Interestingly, the third Sonine coefficient $a_3$ is of the same order of magnitude with respect to the
small parameter $\delta$, as the second Sonine coefficient $a_2$, albeit five times smaller. This
conclusion is in a sharp contrast with the hypothesis suggested in Ref. \cite{huthmann}, that the Sonine
coefficients $a_k$ are of an ascending order of some small parameter $\lambda$, that is, $a_k \sim
\lambda^k$.
\begin{figure}[htbp]
\includegraphics[width=0.7\columnwidth]{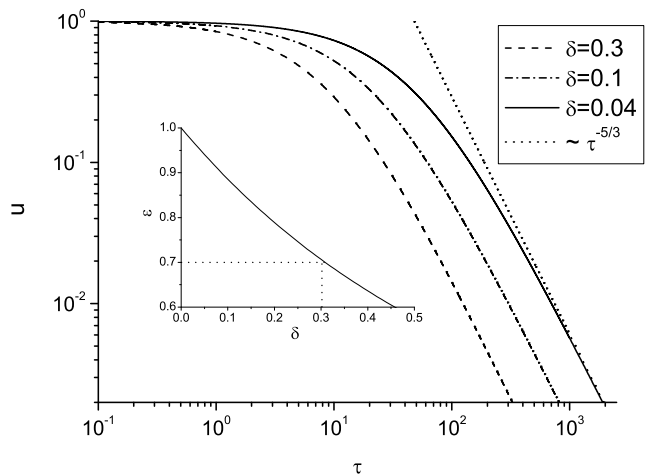}
\caption{Evolution of the dimensionless granular temperature $u(\tau)=T(\tau)/T(0)$, as it follows from
the numerical solution of Eqs. (\ref{system}), for different values of the dissipative parameter
$\delta$. Time is measured in the collision units $\tau_0$ (see the text). For $\tau \gg 1$ all curves
demonstrate the same slope, $u(\tau) \sim \tau^{-5/3}$ (shown by the dotted line), in accordance with
the theoretical prediction, Eq. (\ref{u11}). The inset illustrates the dependence (\ref{eps1}) of the
restitution coefficient $\varepsilon$ on the dissipative parameter $\delta$ for $u=1$ for collisions
with the characteristic thermal velocity, that is for $\left|\vec{c}_{12} \cdot \vec{e}\, \right|=1$.
The particular value of $\varepsilon=0.7$, discussed below,  corresponds to $\delta \simeq 0.3$.}
\label{Gu}
\end{figure}

\begin{figure}[htbp]
\includegraphics[width=0.7\columnwidth]{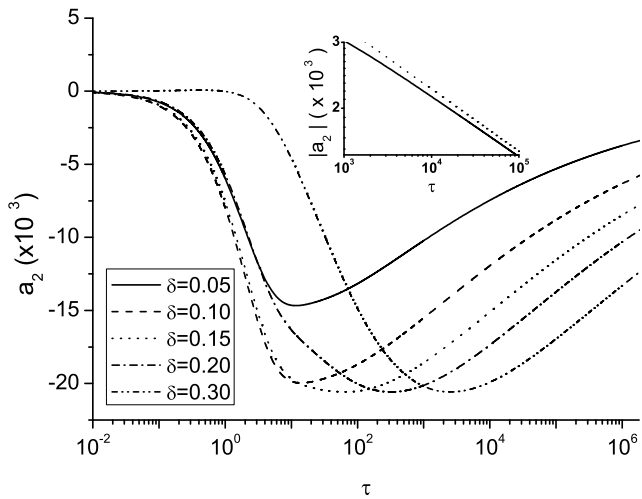}
\caption{Evolution of second Sonine coefficient $a_{2}(\tau)$.  At the first stage of the evolution
$|a_{2}(\tau)|$ increases, then reaches the maximum  value $|a_{2\max}|$, and eventually decays  to
zero. With the increasing dissipation parameter $\delta$ the maximum $|a_{2\max}|$ shifts to later time.
The inset shows the asymptotic dependence of $\left|a_2(\tau)\right|$ (full line) together  with the
analytical result (\ref{eq:a2}) (dotted line)  for $\delta=0.01$ in the log-log scale. } \label{Ga2}
\end{figure}

\begin{figure}[htbp]
\includegraphics[width=0.7\columnwidth]{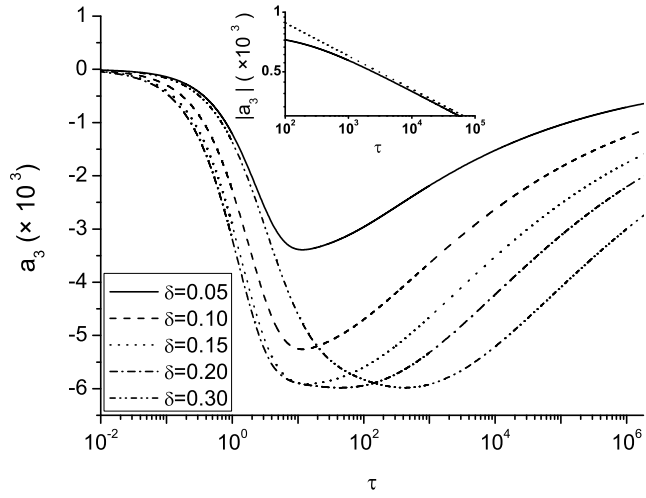}
\caption{Evolution of the third Sonine coefficient $a_{3}(\tau)$. The inset shows the asymptotic
dependence of $\left|a_3(\tau)\right|$ (full line) together with the analytical result (\ref{eq:a3})
(dotted line)  for $\delta=0.01$ in the log-log scale.  } \label{Ga3}
\end{figure}

The numerical solution of Eqs. (\ref{system}) confirms the obtained asymptotic dependence, Eqs.
(\ref{eq:a2})-(\ref{u11}). This is seen in Fig. \ref{Gu}, where the time dependence of the reduced
temperature $u(\tau)$ is plotted and in Figs. \ref{Ga2} and \ref{Ga3} (insets).  Fig. \ref{Gu} demonstrates
that the larger the dissipative parameter $\delta$, the earlier the asymptotic behavior of $u(\tau)$ is
achieved.

The numerical solution for the Sonine coefficients $a_2(\tau)$ and $a_3(\tau)$, shown in Fig. \ref{Ga2}
and \ref{Ga3}, corresponds to the initial Maxwellian distribution,  $a_{2}(0)=a_{3}(0)=0$. As it follows
from the figures, the absolute values of the both coefficients initially  increase, reach their maxima
$\left|a_{2\max}\right|$ and $\left|a_{3\max}\right|$  and, eventually, decay  to zero. In other words,
the velocity distribution function for viscoelastic particles evolves towards the Maxwellian. It is
interesting to note, that the maxima $\left|a_{2\max}\right|$ and $\left|a_{3\max}\right|$ first
increase  with the increasing dissipative parameter $\delta$, then saturate  at $\delta \simeq 0.15$ and
do not anymore grow. The location of the maxima, however, shifts with increasing $\delta$ to later time,
Figs. \ref{Ga2} and \ref{Ga3}. This illustrates the general tendency -- the larger the dissipation
parameter, the slower the gas evolution. Again we see that the third Sonine coefficient $a_3$ is of  the
same order of magnitude as $a_2$ (although a few times smaller) for all evolution stages.

\begin{figure}[htbp]
\includegraphics[width=0.7\columnwidth]{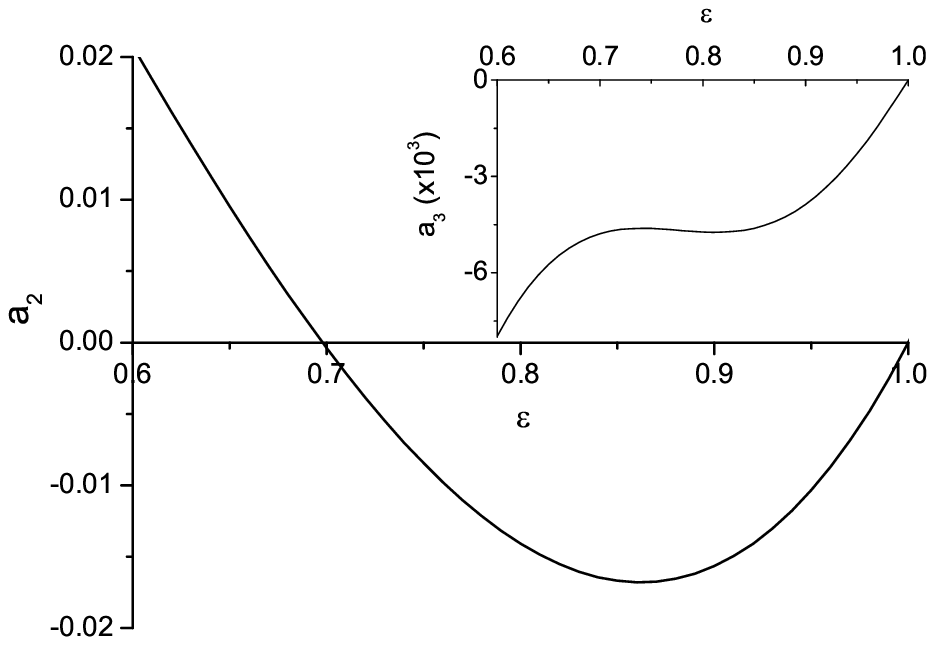}
\caption{Dependence of the second and third (inset) Sonine coefficients on the restitution coefficient
$\varepsilon$ for the case of a constant $\varepsilon$ \cite{book,vannoije,breakdown}. Note that
$\varepsilon \simeq 0.7$ corresponds to the vanishing $a_2$. } \label{Ga2a3eps}
\end{figure}

To understand  the observed behavior of the Sonine coefficients, consider the dependence of these
coefficients on $\varepsilon$ for a constant restitution coefficient (Fig. \ref{Ga2a3eps})  and the
respective dependence of the restitution coefficient $\varepsilon$ on the dissipative parameter $\delta$
(Fig. \ref{Gu}, inset). For $0< \delta \, < 0.15$ [which corresponds to $0.85 < \varepsilon <1$, Fig.
\ref{Gu} (inset)] one observes a fast  relaxation, on a collision time scale, of the Sonine coefficients
$a_2$, $a_3$ to their maximal values $\left|a_{2\max}\right|$, $\left|a_{3\max}\right|$, roughly
corresponding to the respective values for the constant $\varepsilon$. In a course of time the granular
temperature $u(\tau)$ decreases and the effective restitution coefficient alters in accordance with
decreasing dissipative parameter $\delta^{\prime}(\tau) = \delta \, \left( 2 u(\tau) \right)^{1/10}$, that is,
the effective $\varepsilon$ increases with time. For  this interval of $\varepsilon$ ($0.85 <
\varepsilon < 1  $) the increasing restitution coefficient implies the decrease of the absolute values
of the Sonine coefficients, Fig. \ref{Ga2a3eps}, which is indeed observed in the evolution of $a_2(\tau)$
and $a_3(\tau)$. The larger the $\delta$ (i.e.  the smaller the effective restitution coefficient), the
larger  the maxima $\left|a_{2\max}\right|$ and  $\left|a_{3\max}\right|$, in accordance with the
dependencies $a_2(\varepsilon)$ and  $a_3(\varepsilon)$ depicted in  Fig. \ref{Ga2a3eps}.

Similarly, for $0.15< \delta  \, < 0.30$ [which corresponds to $0.7 < \varepsilon <0.85 $, Fig. \ref{Gu}
(inset)] the initial fast relaxation to the related values of $\left|a_{2\max}\right|$ and
$\left|a_{3\max}\right|$ first takes place. Then the decreases of $\delta^{\prime}(\tau)$ and the
respective increase of the effective $\varepsilon$ causes the increase of the absolute value of
$a_2(\tau)$, until it reaches the  maximum, corresponding to $\delta^{\prime}(\tau) =0.15$ (or $\varepsilon
=0.85$), Fig. \ref{Ga2a3eps}. The further decrease of $\delta^{\prime}(\tau)$ leads to the according decay
of $a_2(\tau)$, in agreement with the dependence of $a_2(\tau)$ shown in Fig. \ref{Ga2}. This qualitatively
explains the evolution of $a_2(\tau)$ and the saturation of its maximum $\left|a_{2\max}\right|$ for
$\delta > 0.15$. The qualitative behavior of the third Sonine coefficient may be explained analogously.

\begin{figure}[htbp]
\includegraphics[width=0.7\columnwidth]{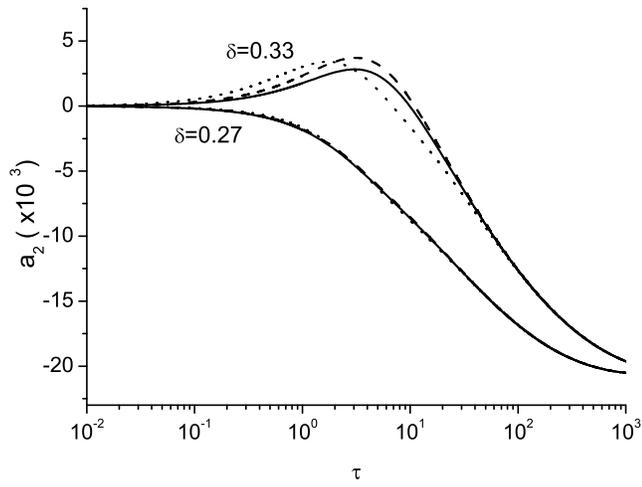}
\caption{The time dependence of the second Sonine coefficient $a_2(\tau)$, computed with a different
number of terms in the expansion (\ref{eps1}) for the restitution coefficient. The full lines correspond
to the accuracy up to ${\cal O}(\delta^{10})$, the dashed lines -- up to ${\cal O}(\delta^{19/2})$ and
the dotted lines -- up to ${\cal O}(\delta^{9})$. Note that for $\delta =0.33$ the accuracy up to ${\cal
O}(\delta^{10})$ is insufficient to obtain a reliable  convergence. } \label{Ga2Odelta}
\end{figure}

It is interesting to note that the observed dependence of $a_2(\tau)$ with the new restitution
coefficient (\ref{eps1}) differs qualitatively for $\delta >0.145$ from that obtained previously for the
old restitution coefficient (\ref{eps}). While in  the latter case a positive bump at initial time,
$\tau \sim 1 $ was detected for $\delta >0.145$ \cite{vdve}, in the former case the positive bump
appears  at much larger $\delta > 0.3$, Fig. \ref{Ga2Odelta}. This is again in agreement with the
behavior, expected from the dependence of $a_2 (\varepsilon)$ for a constant $\varepsilon$: The
coefficient $a_2$ becomes positive for $\varepsilon <0.7$, which corresponds to $\delta > 0.3$, Fig.
\ref{Gu} (inset). Note, however, that with increasing $\delta$ more and more terms in the expansion
(\ref{eps1}) are to be kept. While for $\delta =0.27$ it suffice to keep terms up to $\delta^9$, for
$\delta =0.33$ the Sonine coefficients $a_2$ computed with the accuracy ${\cal O} (\delta^9)$, ${\cal O}
(\delta^{19/2})$ and ${\cal O} (\delta^{10})$ noticeably differ, Fig. \ref{Ga2Odelta}. Therefore we
conclude that the revised restitution coefficient for a  viscoelastic impact (\ref{eps1}) may be
accurately used up to $\delta =0.3$. The loss of the accuracy in the computation of the Sonine
coefficients for larger values of $\delta $ may be a manifestation of the breakdown of the Sonine
polynomials expansion, as it has been found previously for a constant restitution coefficient
\cite{breakdown}. Similarly as in the case of constant $\varepsilon$ \cite{breakdown}, we expect that in
the domain of convergence of the Sonine expansion, $\delta <0.3$, the magnitude of the next-order Sonine
coefficients $a_4$, $a_5,\,  \ldots$ is very small, so that an acceptable accuracy may be achieved with
the use of the two coefficients, $a_2$ and $a_3$ only.

\section{High-velocity tail of the velocity distribution function}

The expansion (\ref{sonine}) refers to the main part of the velocity distribution, $c \sim 1$, that is,
to  the velocities, close to the thermal one,  $v_{T}$. The high-velocity tail  $c \gg  1$ is however
exponentially overpopulated \cite{esipov}. It develops in a course of time, during a first few tens of
collisions \cite{vdeps,tail2007}. For viscoelastic particles the velocity distribution reads for $c \gg
1$  \cite{vdve}:
\begin{equation}
\tilde{f}(c, \tau)\sim\exp\left(-\varphi(\tau) c \right)  \, , \label{exp-1}
\end{equation}
where the function $\varphi(\tau)$ satisfies the equation \cite{vdve},
\begin{equation}
\dot{\varphi}+\frac{1}{3}\mu_{2}B\varphi=\pi B \, , \label{varphi}
\end{equation}
with  $B$ and $\mu_2$ defined previously. Using $\mu_2(\tau) $, obtained by the formula manipulation
program \cite{book} (see the discussion above),  we solve numerically Eq. (\ref{varphi}) to obtain
$\varphi(\tau)$. In the linear approximation, $\mu_{2} \approx 6.49 \, \delta \,
\left(2u(\tau)\right)^{1/10}$,  the function $\varphi(\tau)$ has the form:
\begin{equation}
\varphi(\tau)=(b/\delta) (1+ \tau/\tau_0)^{\frac{1}{6}}
\label{phias}
\end{equation}
with $ b \approx 1.129$ \cite{book}.

Following \cite{vdeps} we neglect the transition region between the main part of the velocity
distribution function, $c \sim 1$, and its high-energy part, $c \gg 1 $ and write the distribution
function as
\begin{eqnarray}
\tilde{f}(c,\tau)&=A(\tau)c^2 \exp(-c^{2})\left(1+a_{2}(\tau)S_{2}(c^{2})+a_{3}(\tau)S_{3}(c^{2})\right)\Theta(c^{*}-c)+\nonumber\\
&+B(\tau)c^2 \exp \left(-\varphi(\tau)c \right)\Theta(c-c^{*}) \label{fct}
\end{eqnarray}
The coefficients $A(\tau)$, $B(\tau)$ and the threshold velocity $c^{*}$, which separates the main and the
tail part of  $\tilde{f}(c,\tau)$ can be obtained, using the normalization condition:
\begin{equation}
\int \tilde{f}(c ) dc =1 \label{norm}
\end{equation}
and the continuity condition for the function itself and its first derivative  \cite{vdeps}:
\begin{eqnarray}
\tilde{f}(c^{*}-0,\tau)=\tilde{f}(c^{*}+0,\tau)\label{ff}\\
\frac{\partial\tilde{f}(c^{*}-0,\tau)}{\partial c}=\frac{\partial\tilde{f}(c^{*}+0,\tau)}{\partial
c}\label{dfdf}
\end{eqnarray}
Substituting Eq. (\ref{fct}) into (\ref{norm}), (\ref{ff}) and (\ref{dfdf}) we arrive at,
\begin{eqnarray}
\cases{(2c^{*}-\varphi)\left(1+a_{2}S_{2}(c^{*2})+a_{3}S_{3}(c^{*2}) \right)=
a_{2}\left(2c^{*3}-5c^{*}\right)+a_{3}
\left(7c^{*3}-c^{*5}-\frac{35}{4}c^{*}\right)\cr
B=A\exp(-c^{*2}+\varphi c^{*})\left(1+a_{2}S_{2}(c^{*2}) +a_{3}S_{3}(c^{*2})\right)\cr
\left(12\sqrt{\pi} {\rm erf}(c^{*})
\exp(c^{*2})-12a_{2}c^{*5}+30a_{2}c^{3*}+35a_{3}c^{*3}-28a_{3}c^{*5}+4a_{3}c^{*7}-24c^{*}\right)\cr
\times\frac{A}{48}\exp(-c^{*2})+\frac{B}{\varphi^{3}}\left(2+\varphi c^{*}(2+\varphi c^{*})\right)\exp(-\varphi c^{*})=1}
\label{system1}
\end{eqnarray}
where $\varphi(\tau)$ is the solution of Eq. (\ref{varphi}).

\begin{figure}[htbp]
\includegraphics[width=0.7\columnwidth]{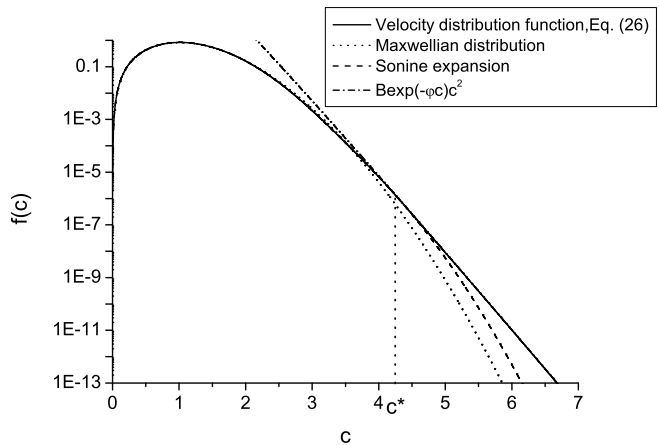}
\caption{A typical velocity distribution function in a granular gas of viscoelastic particles at
$\tau=50$ for $\delta=0.30$ (Eq. (\ref{fct}), solid line), which is represented as a sum of velocity
distribution function, obtained in the Sonine polynomial approximation (Eq. (\ref{eq:sonine3}), dashed
line), and the exponential function (Eq. (\ref{exp-1}), dash-dotted line). For comparison the Maxwellian
distribution is also shown (dotted line). The threshold velocity, $c^{*}=4.31$, may be compared with the
respective quantity  $c^{*}=3.77$ for a gas of particles with a constant restitution coefficient
$\varepsilon \simeq 0.71$.} \label{Gfvd}
\end{figure}

To find the amplitudes $A(\tau)$ and $B(\tau)$ together with the threshold velocity $c^*$ the system
(\ref{system1}) was solved numerically.

The asymptotic dependence of $c^*$ on $\tau$ may be easily found if we take into account that $a_2$ and
$a_3$ are of the same order of magnitude for $\tau \gg 1$, while $\varphi(\tau) \gg 1 $ and $c^* \gg 1$.
Keeping in the first equation in (\ref{system1}) only the largest terms, yields
$(2c^* -\varphi)a_3c^{* \, 6} \simeq -a_3c^{*\,5}$,  
which implies that
\begin{equation}
c^*(\tau) \simeq \varphi(\tau)/2 =(b/2\delta) (1+ \tau/\tau_0)^{1/6}
\label{cas}
\end{equation}

\begin{figure}[htbp]
\includegraphics[width=0.7\columnwidth]{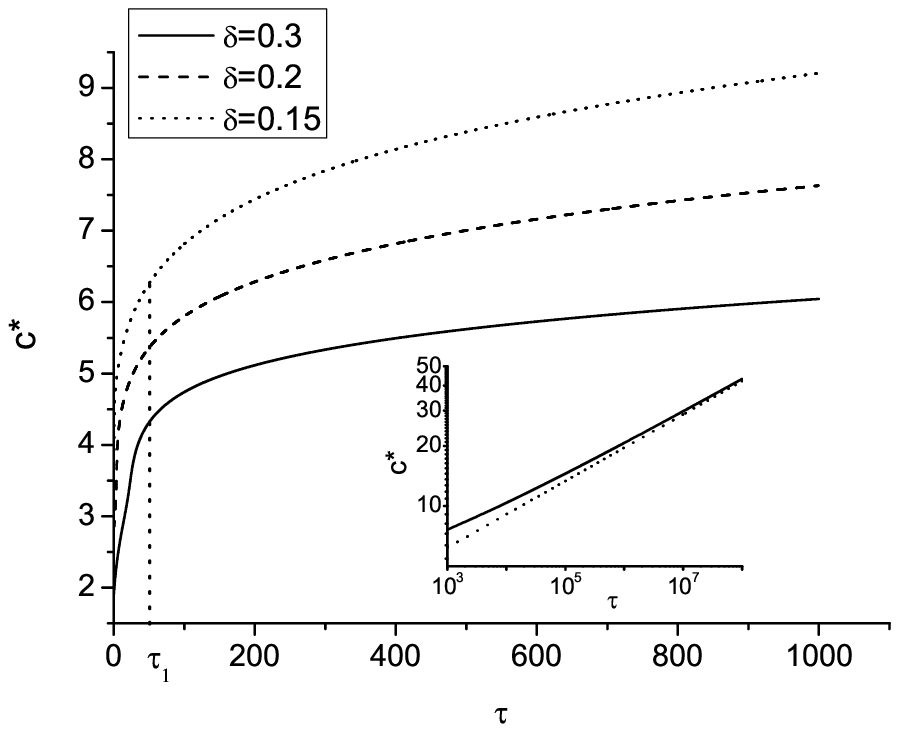}
\caption{Evolution of the threshold velocity $c^{*}$ in a gas of viscoelastic particles for
$\delta=0.3,\,  0.2, \, 0.15 $ (from top to bottom). The corresponding quantities for the case of a
constant restitution coefficient are: $c^{*}(\varepsilon \simeq  0.71)=3.77$, $c^{*}(\varepsilon \simeq
0.79)=3.65$  and  $c^{*}(\varepsilon \simeq 0.84)=3.60$ \cite{vdeps}. In a course of time the threshold
velocity shifts to larger values, that is, the high-energy tail becomes less pronounced. The inset shows
the asymptotic dependence of $c^{*}$ (full line) together with the analytical result (\ref{cas}) (dotted
line) for $\delta=0.2$ in the log-log scale.  } \label{Gc8}
\end{figure}
The typical velocity distribution function, computed for
$\tau=50$ and $\delta=0.3$ is shown in Fig. \ref{Gfvd}. The threshold velocity, $c^*$ separating the
main and the tail part of the velocity distribution reads in this case, $c^{*} \approx 4.31$. The
threshold velocity increases with decreasing inelasticity  $\delta$ and shifts to larger values at later
time, Fig. \ref{Gc8}, which means that the high-energy tail becomes less pronounced.  Again, we see that
in contrast to a gas of particles with a constant restitution coefficient, where the tail of
$\tilde{f}(c)$ persists after its relaxation, in a gas of viscoelastic particles the velocity
distribution function tends to a Maxwellian.

\section{Self-diffusion}
Self-diffusion is the only transport process which takes place in a granular gas in a homogeneous
cooling state: In spite of the lack of macroscopic currents, a current of tagged particles, identical to
the particles of the surrounding gas, but somehow marked, may exist. Moreover, the diffusion coefficient
$D$ is directly related to the mobility coefficient of the tagged particles $\kappa$ via the Einstein
relation, $\kappa \simeq D/T$, which approximately holds true for granular gases, e.g.
\cite{Puglisi2004,Garzo2004}. The mean-square displacement of the tagged particles reads \cite{book}
\begin{equation}
\left< \left[ \Delta r (\tau) \right]^2 \right> = \int^{t} D(t^{\prime}) d t^{\prime} \, ,
\label{eq:delta_r2}
\end{equation}
where the time-dependent diffusion coefficient (diffusivity) is the solution of the following equation
\cite{book}:
\begin{equation}
-\zeta T\frac{\partial D}{\partial T} + D\tau_{v,ad}^{-1}=\frac{T}{m} \, .\label{initial}
\end{equation}
\begin{figure}[htbp]
\includegraphics[width=0.7\columnwidth]{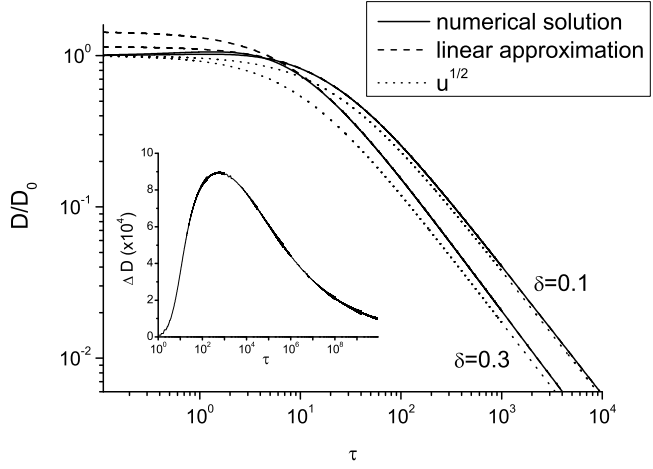}
\caption{ Self--diffusion coefficient as a function of time, measured in the collision units $\tau_0$
(see the text).  The dissipative parameters, from the top to bottom are $\delta=0.3$ and $\delta=0.1$.
$D_0$ is the Enskog self-diffusion coefficient in a gas of elastic particles at the initial temperature
$T_0=T(0)$. The full lines correspond to the complete solution of Eq. (\ref{initial}), while the dashed
lines depict the linear approximation to  $D$,  (\ref{eq:linear_D}). Note that after a short period of
time of about ten collisions, the complete and linear solution become practically indistinguishable.
With the dotted lines we show $u^{1/2}(\tau) = \left( T(\tau)/ T_0 \right)^{1/2}$, which is equal to the
ratio of two diffusion coefficients for elastic particles at temperatures $T(\tau)$ and $T_0$. The
plotted ratio $D(\tau)/D_0$ clearly tends to $u^{1/2}(\tau)$, which manifests that the self-diffusion in
a gas of viscoelastic particles tends to that in a gas of elastic particles. The inset  shows the
relative deviation $\Delta D=(D-\tilde{D})/D$ of the diffusion coefficient $D$ computed with the use of
both Sonine coefficients, $a_2$ and $a_3$  from the respective value, $\tilde{D}$, obtained with the use
of $a_2$ only ($\delta=0.3$). Naturally, the location of maximum of $\Delta D$ coincides with that of $\left|a_3
(t)\right|$, Fig.  \ref{Ga3}. } \label{Gdiff}
\end{figure}
Here $\zeta=(2/3)B \mu_2 $ is, as previously,  the cooling coefficient and $\tau_{v,ad}$ is the velocity
correlation time, which characterizes the time after which the memory about the initial particle
velocity is lost; it reads in terms of the distribution function \cite{book}:
\begin{eqnarray}
\label{eq:tau_v}
&&\tau_{v,ad}^{-1} = \frac16 v_T(t)g_2(\sigma) \sigma^2 n \times \\
&& ~~~~~~~~~ \times \int d\vec{c}_1\vec{c}_2  \int d \vec{e} \, \Theta(-\vec{c}_{12} \cdot \vec{e} \, )
 \left|\vec{c}_{12} \cdot \vec{e} \, \right| \tilde{f}(\vec{c}_1,t ) \tilde{f}(\vec{c}_2,t )
 (1-\varepsilon) (\vec{c}_{12} \cdot \vec{e}\, )^2 \nonumber \,.
\end{eqnarray}
With the approximation (\ref{eq:sonine3}) for $ \tilde{f}(\vec{c},t )$ $\tau_{v,ad}$ was calculated up
to $\delta^{10}$, using the formula manipulation program, described in \cite{book}. Here we present for
illustration only its linear with respect to $a_2$, $a_3$ and $\delta^{\prime}$ part:
\begin{equation}
\frac{\tau_{v,ad}}{\tau_{E}(0)}=\left( 1+\frac{3}{16}a_2+\frac{1}{64}a_3 \right)
-\sqrt{2\pi}\omega_0\left(\frac{1}{8}+\frac{3}{100}a_2 +\frac{1}{500} a_3\right)\delta^{\prime}(\tau)
\end{equation}
Here $\tau_{E}(0)$ is the Enskog velocity correlation time of elastic particles at initial time
$\tau=0$:
\begin{equation}
\label{eq:Enskog_time} \tau_{E}^{-1}(0)=\frac{8}{3}\sqrt{\frac{\pi T(0)}{m}}n\sigma^2g_2 \, .
\end{equation}
Using the obtained expressions for  $\mu_2 $ and $\tau_{v,ad}$ up to ${\cal O}(\delta^{10})$, we solve
Eq. (\ref{initial}) numerically and compute the diffusion coefficient $D(\tau)$. In Fig. \ref{Gdiff} the
ratio of $D(\tau)/D_0$ is plotted,  where
\begin{equation}
\label{eq:D0} D_0 = \tau_{E}(0)T(0)/m \, ,
\end{equation}
is the Enskog self-diffusion coefficient  for a gas of elastic particles at initial temperature $T(0)$.
As it is clearly seen from the figure, the diffusion coefficient decreases with time in a way, that the
ratio  $D(\tau)/D_0$ approaches $u^{1/2}(\tau)= \left(T(\tau)/T(0) \right)^{1/2}$ -- the ratio of the two
diffusion coefficients for elastic particles at the current temperature $T(\tau)$ and the initial
temperature $T(0)$. Hence in a course of time the self-diffusion in a gas of viscoelastic particles tends
to that in a gas of elastic particles.

In the linear approximation with respect to the small dissipative parameter $\delta$ one can obtain an
analytical expression for $D(\tau)$. Keeping only first-order terms in the expressions for $\mu_2$, $a_2$,
$a_3$ and $\tau_{v,ad}$ and substituting them into Eq. (\ref{initial}) yields the diffusion coefficient:
\begin{equation}
\label{eq:linear_D} D=D_{0}u^{1/2}(\tau) \left(1+\frac{4239}{16000}\, \omega_0\,
\sqrt{\frac{2}{\pi}}\, \delta^{\prime} \right) \, ,
\end{equation}
where, as previously, $\delta^{\prime} = \delta \left( 2 u(\tau) \right)^{1/10}$ and $\omega_0 \approx
6.485$. Note that due to the account of the third Sonine coefficient in $\mu_2$ and $\tau_{v,ad}$ the
linear approximation (\ref{eq:linear_D}) for $D(\tau)$ differs from the previously obtained result
\cite{book}.

The linear approximation (\ref{eq:linear_D}) is compared in Fig. \ref{Gdiff} with the complete numerical
solution. As it follows from the figure, after about ten collision per particle the two solution become
practically indistinguishable. As it may be seen from the inset of  Fig. \ref{Gdiff}, the impact of the
third Sonine coefficient on the behavior of self-diffusion coefficient is rather small even for the large value of dissipation parameter $\delta =0.3$.

\section{Conclusion}

We study  evolution of a granular gas of viscoelastic particles in a homogeneous cooling state. We use a
new expression for the restitution coefficient $\varepsilon$, which accounts for the delayed recovery of
the particle material at a collision and allows to model collisions with much larger dissipation,  as
compared to previously available result for $\varepsilon$. We analyze the velocity distribution function
and the self-diffusion coefficient. To describe the deviation of the velocity distribution function from
the Maxwellian we use the Sonine polynomial expansion. In contrast to the commonly used approximation,
which neglects all terms in the Sonine expansion beyond the second one, we consider explicitly the third
Sonine coefficient. We detect a complicated  evolution of this coefficient and observe that it is of the
same order of magnitude, with respect to the (small) dissipative parameter $\delta$, as the second
coefficient. This contradicts the existing hypothesis \cite{huthmann}, that the subsequent Sonine
coefficients $a_2$, $a_3 \ldots$, $a_k$ are of an  ascending order of some small parameter,
characterizing particles inelasticity. Similarly as for the case of a constant restitution coefficient,
we obtain an indication of divergence of the Sonine expansion for large dissipation, $\delta >0.3$.  For
the asymptotic long-time behavior we derive  analytical expressions for both Sonine coefficients, which
agree well with the numerical data.

Using the obtained third Sonine coefficient we compute the self-diffusion coefficient $D$ and derive an
analytical expression for $D$ in the linear, with respect to $\delta$, approximation. We show that the
complete solution approaches the approximate analytical solution after a transient time of  about ten
collisions per particle. We observe, that in spite of the importance of $a_3$ for an accurate
description of the velocity distribution function, its impact on $D$ is rather small.

 We also study the evolution of the high-energy tail of the
velocity distribution. Using the equation for the time-dependent slope of the tail and the obtained
Sonine coefficients we find the amplitude of the tail and the threshold velocity, which demarcates  the
main part of the velocity distribution and the high-energy tail. We find the analytical expression for
the asymptotic behavior of the threshold velocity, which agrees well with numerics,  and observe, that
in a course of time it shifts to larger values; this implies that the high-velocity tail becomes less
pronounced.

Such behavior of the threshold velocity, of the Sonine coefficients and of the coefficient of
self-diffusion, naturally, manifests that the properties of the system tend to those of a gas of elastic
particles; our theory quantifies evolution towards this limit.

\end{document}